# PHD-Store: An Adaptive SPARQL Engine with Dynamic Partitioning for Distributed RDF Repositories

Razen Al-Harbi

KAUST, Saudi Arabia
razen.harbi@kaust.edu.sa

Yasser Ebrahim

EPFL, Switzerland
yasser.ibrahim@epfl.ch

Panos Kalnis

KAUST, Saudi Arabia
panos.kalnis@kaust.edu.sa

## ABSTRACT

Many repositories utilize the versatile RDF model to publish data. Repositories are typically distributed and geographically remote, but data are interconnected (e.g., the Semantic Web) and queried globally by a language such as SPARQL. Due to the network cost and the nature of the queries, the execution time can be prohibitively high. Current solutions attempt to minimize the network cost by redistributing all data in a preprocessing phase, but there are two drawbacks: ($i$) redistribution is based on heuristics that may not benefit many of the future queries; and ($ii$) the preprocessing phase is very expensive even for moderate size datasets.

In this paper we propose PHD-Store, a SPARQL engine for distributed RDF repositories. Our system does not assume any particular initial data placement and does not require prepartitioning; hence, it minimizes the startup cost. Initially, PHD-Store answers queries using a potentially slow distributed semi-join algorithm, but adapts dynamically to the query load by incrementally redistributing frequently accessed data. Redistribution is done in a way that future queries can benefit from fast hash-based parallel execution. Our experiments with synthetic and real data verify that PHD-Store scales to very large datasets; many repositories; converges to comparable or better quality of partitioning than existing methods; and executes large query loads 1 to 2 orders of magnitude faster than our competitors.

## 1. INTRODUCTION

RDF [3] datasets consist of ⟨subject, $predicate$, object⟩ triples, where the predicate represents a relationship between two entities: the subject and the object. They can be viewed as directed labeled graphs, where vertices and edge labels correspond to entities and predicates, respectively. Figure 1 shows an example RDF graph of students and professors in an academic network. The RDF model does not require a predefined schema and is a versatile way to represent information from diverse sources. It is used in the Semantic Web and in a variety of applications including social networks, online shopping, scientific databases, etc.

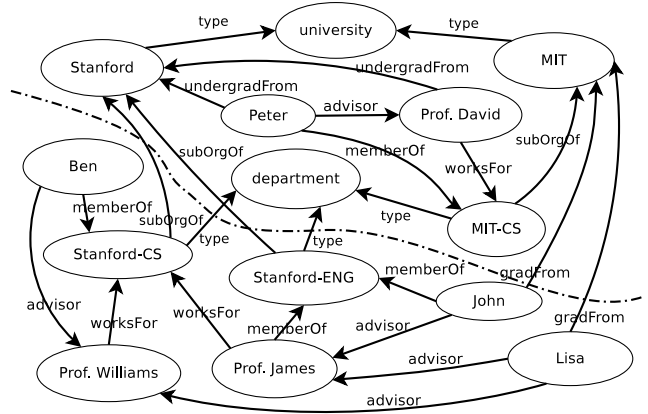

Figure 1: Example RDF graph. An edge with its associated vertices corresponds to an RDF triple; e.g., ⟨Prof.Williams, $worksFor$, Stanford-CS⟩. Dotted line depicts a MinCut partitioning.

SPARQL is the standard query language for RDF. Queries consist of a set of RDF triple patterns, where some of the columns are variables. For example, let $Q_{prof}$ be:

```
SELECT ?x WHERE {
    ?x      worksFor  Stanford-CS
    Lisa    advisor   ?x }
```

$Q_{prof}$ returns Lisa's advisors who work for Stanford CS. The query corresponds to the graph of Figure 2(a). The answer is the set of bindings of ?x that render the query graph isomorphic to subgraphs in the data. In our example, ?x ∈ {Prof.Williams, Prof.James} (see Figure 1).

Let the data be stored in a table $D(s,p,o)$, where rows are RDF triples ⟨s, p, o⟩. To answer $Q_{prof}$, first decompose it into two subqueries and answer them independently by scanning table $D$: $q_1 \equiv \sigma_{p=\text{worksFor} \wedge o=\text{Stanford-CS}}(D)$ and $q_2 \equiv \sigma_{s=\text{Lisa} \wedge p=\text{advisor}}(D)$. Then, join the intermediate results on the subject and object attribute: $q_1 \bowtie_{q_1.s=q_2.o} q_2$. If all data are on the same server, the plan can be executed efficiently by a system like RDF-3X [25], which indexes all combinations of the three attributes of $D$.

In many practical applications $D$ is distributed among many geographically remote repositories. For instance, our example involves two universities, Stanford and MIT. It is natural to allow each university to handle its own data.

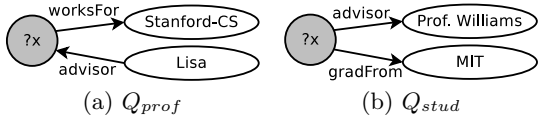

(a) $Q_{prof}$  (b) $Q_{stud}$

**Figure 2:** SPARQL queries: (a) Find Lisa's advisors who work for Stanford CS. (b) Find the advisees of Prof. Williams who graduated from MIT.

Therefore, triple ⟨Lisa, $advisor$, Prof.Williams⟩ is expected to be in Stanford's server, whereas MIT will store ⟨Lisa, $gradFrom$, MIT⟩. Consider query $Q_{stud}$ = "Find the students who graduated from MIT and are advised by Prof. Williams", depicted in Figure 2(b). None of the servers has all necessary triples to execute the join. Therefore, intermediate results must be transferred from Stanford to MIT, or vice-versa. In practice, the intermediate results can be very large and the bandwidth between remote servers may fluctuate considerably. Due to the communication cost, the resulting response time can be unacceptably long.

A possible solution could be based on each server caching its partial results. If the same query is asked again, each server returns its cached results to the master, which consolidates them into the final answer. However, this approach suffers from the following drawbacks: ($i$) It only works well if each query is asked many times. ($ii$) A cached query cannot answer another query that has the same pattern but different variables.

Another approach to minimize the communication cost is to prepartition the data. Previous work [15, 18, 27] introduced a preprocessing step where the entire dataset is partitioned among the servers by hashing on the subject or object column. In our example, let subject be the hash key, forcing triples ⟨Lisa, $advisor$, Prof.Williams⟩ and ⟨Lisa, $gradFrom$, MIT⟩ to move to the same server. Then, $Q_{stud}$ can be answered without communication. This is also true for any star-shaped query where subject is the variable. Unfortunately, because of the different hash keys, ⟨Lisa, $advisor$, Prof.Williams⟩ and ⟨Prof.Williams, $worksFor$, Stanford-CS⟩ are likely to be in different servers; therefore, query $Q_{prof}$ still requires considerable communication.

A recent work [20] employs a MinCut [22] algorithm to partition the graph during preprocessing. This step generates partitions with roughly balanced number of vertices and minimal number of edges between different partitions. The dotted line in Figure 1 depicts such a partitioning. Observe that all necessary triples for $Q_{prof}$ are now in the lower partition; therefore, the query can be executed without communication. This is true for $Q_{stud}$ as well, assuming that triples follow the placement of their subject vertex (e.g., ⟨Lisa, $gradFrom$, MIT⟩ will be placed at the lower partition, because of Lisa). Nevertheless, there still exist numerous queries that cross partition boundaries and require significant communication, such as: ⟨?x, $subOrgOf$, ?y⟩ AND ⟨?y, $type$, University⟩.

Current partitioning approaches have three drawbacks: ($i$) Static partitioning (hash-based or MinCut) is not necessarily a good fit for many queries. As explained above, there exist workloads that access many servers no matter how partitioning is done. Hash partitioning, in particular, is useful only for star-shaped queries. ($ii$) The preprocessing step is expensive. MinCut partitioning, in particular, requires to transfer the entire dataset to a central location; run MinCut, which needs several hours on many CPUs and hundreds of GB of RAM even for moderate size graphs; and transfer the resulting partitions back to the servers. ($iii$) The partitioning cost is paid for the entire dataset, even if future queries will access only a small subset of the graph.

We propose PHD-Store, a SPARQL engine for distributed, geographically remote RDF repositories. PHD-Store optimizes the execution of distributed joins without relying on static partitioning. Instead, it adapts by dynamically redistributing portions of the graph that are accessed by the query load. Consider again $Q_{prof}$ in Figure 2(a). Our system selects a vertex in the query graph as a core vertex, say ?x. PHD-Store places the bindings of ⟨?x, $worksFor$, Stanford-CS⟩ using the values of the core (?x) as hash keys. In this case, ⟨Prof.Williams, $worksFor$, Stanford-CS⟩ and ⟨Prof.James, $worksFor$, Stanford-CS⟩ are placed in two different servers (since the keys are different). ⟨Lisa, $advisor$, Prof.Williams⟩ and ⟨Lisa, $advisor$, Prof.James⟩ follow the placement of Prof. Williams and Prof. James, respectively. No further triples are moved. In general, vertices that bind to the core will have their neighbors copied to the same server, in a recursive fashion that results in a tree-shaped distribution. We call this Propagating Hash Distribution (*PHD*). Our method achieves two goals: ($i$) *Minimal communication*: $Q_{prof}$ can now be executed without exchanging intermediate results between the servers; and ($ii$) *Parallel mode*: the required data are redistributed in multiple servers that can work in parallel to minimize response time.

PHD-Store maintains an index of patterns that have been redistributed, and a query optimizer that combines parts of multiple previously redistributed queries to answer a new query in parallel mode. It also accepts any initial placement of data (including random) and starts processing queries immediately. By avoiding the upfront cost and adopting a pay-as-you-go approach, our system can execute tens of thousands of queries within the time it takes our competitors to partition even a small graph. More importantly, the quality of the PHD partitioning is, in general, better, since it is guided by the actual query load. Therefore, PHD-Store scales to very large graphs and many RDF repositories. In summary, our contributions are:

- We introduce PHD-Store, a SPARQL engine for distributed RDF repositories, that does not require expensive preprocessing.

- We propose PHD, an adaptive technique that redistributes data dynamically, in a way that future queries can be executed in parallel mode.

- We evaluate our system using synthetic and real data on a cluster of 21 machines. PHD-Store is initialized in around one minute, whereas our competitors need up to 22 hours. Consequently, PHD-Store can execute a large workload 1 to 2 orders of magnitude faster than existing approaches.

The rest of this paper is organized as follows: Section 2 discusses the related work. Section 3 presents the architecture of PHD-Store. Section 4 discusses the adaptive indexing mechanism, whereas Section 5 explains the query indexing and distributed data management. Section 6 discusses how updates are managed in PHD-Store. Section 7 contains the experimental results and Section 8 concludes the paper.

## 2. RELATED WORK

**Centralized RDF stores.** Early approaches such as RDF-Suite [6] and Sesame [10], store RDF triples ⟨s, p, o⟩ as large tables in a relational database, usually with indices on all three columns. Jena [30] adds support for rich features, such as inference. Abadi et al. [4] use a collection of smaller tables, one for each distinct predicate value and employ a column-based DBMS. More recent systems, such as YARS2 [18] and HPRD [23], implement native RDF stores with specialized indexing. Both, however, lack an efficient query optimizer. RDF-3X [25] is one of the most promising native RDF stores. It maintains indices that cover all permutations of the three columns of RDF triples, uses rigorous byte-level compression and its query optimizer favors fast sort-merge joins. Similar ideas are implemented in Hexastore [29].

**Materialized views.** Recent works attempt to speed up the execution of SPARQL queries by selecting a set of views to materialize based on a given workload [11, 17]; or by materializing a set of path expressions based again on the workload [14]; and introducing query rewriting techniques that use the materialized views [12, 17]. In our approach, we do not generate materialized views or perform any query rewriting. Instead, we redistribute and possibly replicate the data accessed by queries in a way that these queries can be executed in parallel mode. We also introduce a mechanism for indexing queries and managing the redistributed data. Nevertheless, because of replication, we share with materialized views the consistency maintenance problem.

**Hash-based distributed RDF stores.** All existing distributed solutions need to prepartition the data in order to minimize communication during query execution. Several systems [15, 18, 27] use hashing to distribute all data during a preprocessing step. As explained previously, hash partitioning works well only for 1-hop star-shaped queries. Servers work in parallel on their local data without exchanging intermediate results (i.e., no communication and no delay due to synchronization barriers); and on average load is balanced among servers. Unfortunately, hash partitioning is inefficient for queries that are not star-shaped or span more than one hop; such queries are common in SPARQL. Our work extends the idea of hash partitioning to support parallel execution of complex queries.

**Optimized partitioning and replication.** Recent systems partition the graph by applying a minimum cut algorithm, such as METIS [22]. Intuitively, if fewer edges cross partitions, the probability of answering a query without communication among servers is increased. There are two drawbacks: (i) min-cut is extremely expensive; and (ii) as discussed in Section 1, there are still a lot of queries that require communication. Huang et al. [20] remove the high-degree vertices prior to partitioning to reduce the complexity of min-cut. They also enforce the so-called $k$-hop guarantee: vertices are replicated among partitions, such that any query with radius $k$ or less (recall that queries are represented as graphs) can be executed without communication. Unfortunately, partitioning still takes several hours even for moderate size graphs. Moreover, replication increases exponentially with $k$; therefore $k$ must be kept small (e.g., $k \leq 2$ in the experiments of Huang et al). If the radius of a query exceeds $k$, or the query splits around a high-degree vertex[1] (both cases are common in SPARQL), then the query is answered by a series of Map-Reduce jobs.

There also exist relevant systems that focus on data models other than RDF. Schism [13] deals with the problem of data placement for distributed OLTP RDBMS. Using a sample of the workload, Schism minimizes the number of distributed transactions by populating a graph of co-accessed tuples. The graph is partitioned by METIS and data are redistributed to servers accordingly. Tuples accessed in the same transaction are put in the same server. This is not appropriate for SPARQL queries because some queries access large parts of the data that would overwhelm a single machine. Instead, PHD-Store exploits parallelism by executing such a query across all machines in parallel without communication. H-Store [28] is an in-memory distributed OLTP RDBMS that uses a data partitioning technique similar to ours and has been extended [26] to handle skewness in the data and workloads. Nevertheless, H-Store assumes that the schema and complete workload is specified in advance and assumes no ad-hoc queries. Although, these are valid assumptions for OLTP databases, they are not for RDF data stores. NuoDB [2] is a commercial, ACID compliant database that supports SQL. NuoDB does not employ sharding for partitioning the database; rather, it smartly caches atoms in multiple servers based on the workload.

Another recent work by Yang et al. [31] focuses on general graphs but can be applied to RDF data. The entire graph is replicated several times and each replica is partitioned in a different way. Each replica runs an instance of Pregel [24]. The query optimizer directs each query to the most suitable replica that minimizes communication. The method has two drawbacks: (i) there is excessive replication; and (ii) data must be localized in order to build the adjacency list of each vertex, required by Pregel; which is a very inefficient process.

**Eventual indexing.** Idreos et al. [21] introduce the concept of reducing the data-to-query time for relational data. They avoid building indices during data loading; instead, they reorder tuples incrementally based on the query ranges. A recent work [5] focuses on executing queries on raw files, and similarly builds incrementally an index for future queries by amortizing the cost among past file accesses. In PHD-Store, we extend the concept of eventual indexing to dynamic and adaptive graph partitioning. In our problem, graph prepartitioning is very expensive, hence, the potential benefits of eliminating the data-to-query time are large.

## 3. SYSTEM ARCHITECTURE

PHD-Store organizes a large number of independent and geographically remote RDF repositories into a federation, allowing users to pose queries over the union of the entire collection. The system architecture is depicted in Figure 3.

### 3.1 Master and Workers

**Master.** The master node receives queries from the users, generates an execution plan, coordinates the workers, collects the final result and returns it to the user. The master contains a *query index* that stores information about the query patterns that have been redistributed. The query index is used by the query planner to decide whether a new

---

[1] To minimize replication, high-degree vertices utilize 1-hop instead of $k$-hop guarantee.

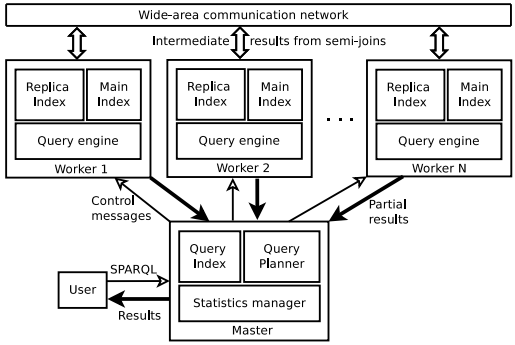

Figure 3: System architecture: Workers correspond to independent RDF repositories.

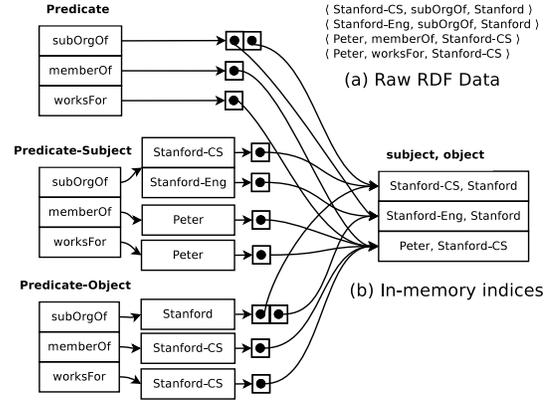

Figure 4: Structure of the main index of a worker: (a) Raw local RDF data at the worker, (b) predicate, predicate-subject and predicate-object indices.

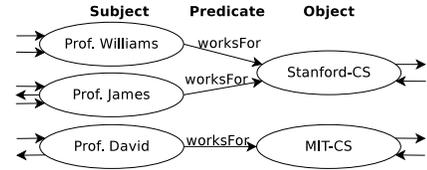

Figure 5: Statistics calculation for $p = \mathtt{worksFor}$, based on the graph of Figure 1. $\overline{p_S} = (3+4+3)/3 = 3.33$; $\overline{p_O} = (4+3)/2 = 3.5$

query can be executed in parallel using the current distribution. The query index will be explained in Section 5. The master also contains a statistics manager that maintains useful statistics about the RDF graph.

**Worker.** Each RDF repository is called a worker. Let there be $N$ workers in the system: $w_1, w_2, \ldots, w_N$. Each worker $w_i$ stores locally a set $D_i$ of triples. The entire dataset $D$ is defined as: $D = \bigcup_1^N D_i$.

Each worker stores its local set of triples in an in-memory structure, called *main index*. It contains a single storage module that consists of the following indices: (*i*) Predicate: given a predicate $p$, return a list of all $\langle\mathtt{subject},\mathtt{object}\rangle$ pairs. (*ii*) Predicate-Subject: given a predicate $p$ and a subject $s$, return a list of all $\langle s, \mathtt{object}\rangle$ pairs. (*iii*) Predicate-Object: given a predicate $p$ and an object $o$, return a list of all $\langle\mathtt{subject}, o\rangle$ pairs. The predicate index is implemented as hash map and the other two as nested hash maps. Typically, the number of predicates in RDF datasets is small compared to the number of triples. To eliminate redundant repetitions of the predicate, the indices store $\langle\mathtt{subject},\mathtt{object}\rangle$ pairs instead of RDF triples. Each pair is stored only once, and only pointers to these pairs are stored in the indices. Figure 4 shows an example. The main index is used to answer queries that cannot be executed in parallel mode.

Each worker also contains an in-memory *replica index* that stores copies of subsets of the entire dataset $D$. The replica index is used when answering queries in parallel mode. Initially, this index contains no data, but it is maintained and updated dynamically by the PHD redistribution process, which may replicate in worker $w_i$ triples from several workers. The process will be explained in Section 4.

### 3.2 Statistics Collection

PHD-Store uses the degrees of vertices in the data to plan for query execution and redistribution, as we explain in Section 4.1. Keeping this information for each vertex in the entire data, is prohibitively expensive. PHD-Store solves the problem by focusing on predicates rather than vertices. For each unique predicate $p$, we calculate the corresponding subject and object scores, defined as follows:

DEFINITION 1 (PREDICATE SCORES). *Let $p$ be a predicate. (*i*) The subject score of $p$, denoted $\overline{p_S}$ is the average degree of all vertices $s$, such that $\langle\mathtt{s}, p, \mathtt{?x}\rangle \in D$. (*ii*) The object score of $p$, denoted $\overline{p_O}$ is the average degree of all vertices $o$, such that $\langle\mathtt{?x}, p, \mathtt{o}\rangle \in D$.*

Figure 5 shows an example for predicate worksFor. $\overline{p_O} = (4 + 3)/2 = 3.5$ because worksFor appears with two unique objects: Stanford CS, whose degree (i.e., in-degree plus out-degree) is 4; and MIT CS, whose degree is 3.

Because $D$ is split in many workers, vertices are typically replicated. Identifying unique vertices, in order to calculate predicate scores, is very expensive due to the communication cost. Therefore, PHD-Store maintains only approximate statistics. Each worker calculates independently its predicate scores and sends them to the master. The master calculates the global scores as the average of the local ones. The process is fast and provides an adequate approximation for query optimization and adaptivity purposes.

### 3.3 Workflow

PHD-Store does not require global indexing and can start from any initial data partitioning, including random. To start a federation of RDF repositories, each worker builds independently its main index and collects statistics. The process is very efficient since it involves a single scan of the local data. Afterwards, PHD-Store can start answering queries immediately.

A user submits a SPARQL query $Q$ to the master. The query planner at the master consults the query index to decide whether $Q$ is processable in parallel mode, or if distributed semi-joins must be used:

**Distributed mode (semi-joins).** If $Q$ cannot be answered in parallel mode, PHD-Store executes the query by a series of distributed semi-joins; the process is described in Algorithm 1. Any worker may contain relevant data. Each worker $w$ sends to all other workers a projection on the join

**Algorithm 1**: Distributed semi-join on $N$ workers. Each worker executes this algorithm

**Input**: Query $Q$ consisting of subqueries $\{q_1, q_2\}$
**Result**: Answer of query $Q$

1 Let $q_1$ and $q_2$ be joined on subject $s$ and object $o$, respectively
2 $RS_1 \leftarrow$ answerSubquery$(q_1)$;
3 $RS_2 \leftarrow$ answerSubquery$(q_2)$;
4 $RS_1[s] \leftarrow \pi_s(RS_1)$; // projection on $s$
5 Send $RS_1[s]$ to all workers;
6 **foreach** worker $w$, $w : 1 \rightarrow N$ **do**
7     Let $RS_{1w}[s]$ denote the $RS_1[s]$ received from $w$
8     Let $CRS_{2w}$ be the candidate triples of $RS_2$ that join with $RS_{1w}[s]$
9     $CRS_{2w} \leftarrow RS_2 \bowtie_{RS_2.o=RS_{1w}[s].s} RS_{1w}[s]$;
10     Send $CRS_{2w}$ to worker $w$;
11     Let $RS_{2w}$ be the $CRS_{2w}$ received from worker $w$
12     Let $RES_w$ be the result after joining with worker $w$
13     $RES_w \leftarrow RS_1 \bowtie_{RS_1.s=RS_{2w}.o} RS_{2w}$;
14 $q_1 \bowtie q_2 \leftarrow RES_1 \cup RES_2 \cup .... \cup RES_N$;
15 send the partial result $q_1 \bowtie q_2$ to master;

column of the relevant triples (line 5). All workers perform the semi-join on the received data (line 9) and send the results back to $w$ (line 10). $w$ finalizes the join (line 13) and returns the partial answer to the master, which forwards it to the user without further processing. Lines 9 and 13 are implemented as local hash-joins, using the main index in each worker. Since $Q$ may consist of multiple subqueries, say $\{q_1, q_2, q_3\}$, the query is evaluated by joining $q_1$ and $q_2$, then joining the result with $q_3$; each join uses Algorithm 1. Note that the master is not involved in the distributed join, and all communication is done among workers.

Since our data are memory resident, each machine uses hash joins as they prove to be competitive to more sophisticated methods [8]. The hash join consists of two phases: build and probe. Our data are already hash-indexed so we do not need the build phase; therefore, the optimizer tries to minimize the number of probes. Currently, the optimizer generates a right-deep join tree, starting with the subquery with the least cardinality. More sophisticated methods like the one discussed in [9] are orthogonal to our work.

**Parallel mode.** In this case, $Q$ is executed in parallel without communication. The master broadcasts $Q$ to all workers. Each worker uses its replica index to construct a partial result, which is sent to the master; no communication among workers is needed. Finally, the master forwards the partial results to the user; no processing is required at the master. Locally, each worker uses hash joins, as discussed above.

**Dynamic redistribution.** PHD-Store monitors the frequency of each query. When the frequency of a query increases above a system-wide threshold, the redistribution process for that query is triggered. After redistribution, the query can be executed in parallel mode. Note that redistribution does not benefit only the query that triggered it; future queries can utilize a subset of the past redistributed data, and can run in parallel mode.

## 4. PHD-STORE ADAPTIVITY

The dynamic redistribution model of PHD-Store is a combination of hash partitioning and $k$-hop replication; however, it is guided by the query load rather than the data itself. Specifically, given a frequent query $Q$, our system selects a special vertex in the query graph called the *core* vertex. The system groups the data accessed by the query around the bindings of this core vertex. To do so, the system decomposes the query into a redistribution tree rooted at the core. Then, starting from the core vertex, first hop triples are hash distributed based on the core bindings. Next, triples that bind to the second level subqueries are collocated and so on. A redistributed query can be executed in parallel without communication. Moreover, queries that have not been redistributed can combine data from already redistributed queries and then run in parallel mode.

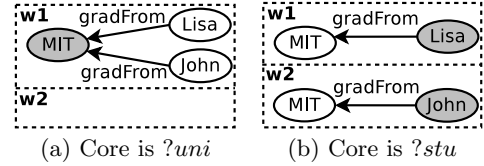

(a) Core is ?*uni*     (b) Core is ?*stu*

**Figure 6: Effect of choice of core on replication. In (a) there is no replication. In (b) MIT is replicated in workers $w_1$ and $w_2$.**

### 4.1 Core Vertex Selection

The choice of the core has a significant impact on the amount of replicated data as well as on the query execution performance. Consider query $Q_1 = \langle \text{?stu}, gradFrom, \text{?uni} \rangle$. Assume there are two workers, $w_1$ and $w_2$, and refer to the graph of Figure 1; MIT is the only university satisfying the query. If ?*uni* is the core, then MIT is hashed to $w_1$. Lisa and John are also propagated to $w_1$ (see Figure 6(a)); therefore there is no replication. On the other hand, if ?*stu* is the core, then Lisa and John are hashed in $w_1$ and $w_2$, respectively. Then MIT is propagated to the core, therefore there are replicas in both workers (see Figure 6(b)). The problem becomes more pronounced when a query has more hops. Consider $Q_2 = Q_1$ AND $\langle \text{?dept}, subOrgOf, \text{?uni} \rangle$ and choose ?*stu* as core. Because MIT is replicated, all the sub-organizations of MIT will also be replicated. This is a significant cost because replication cost grows exponentially with the number of hops [20].

Intuitively, if random walks start from two random vertices (e.g., students), the probability to reach the same well-connected vertex (e.g., university) within a few hops is higher than reaching the same student from two universities. In order to minimize replication, we must avoid reaching the same vertex when starting from the core. Therefore, it is reasonable to select a well-connected vertex as the core. In the literature there are many definitions of what constitutes a well-connected vertex, many of which are based on complex data mining algorithms. In contrast, we employ a definition that poses minimal computational overhead: we assume that well-connectivity is proportional to the degree (i.e., in-degree plus out-degree) of the vertex.

Nonetheless, many RDF datasets follow the power-law distribution in which few vertices are of extremely high degrees. Treating such vertices as cores is problematic because they cause many vertices to be placed in the same worker. Vertices that appear as objects in triples with *rdf:type* predicate are also problematic. Selecting these vertices to be cores will cause all vertices of the same type to be placed in one worker. This would overwhelm the worker and would not take advantage of parallelism [20].

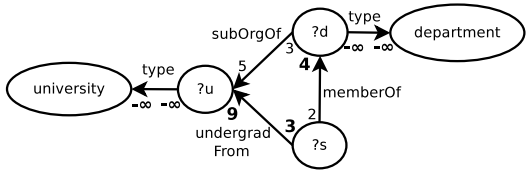

**Figure 7:** Example of vertex score: numbers correspond to $\overline{p_S}$ and $\overline{p_O}$ values. Assigned vertex scores $\overline{v}$ are shown in bold.

Recall from Section 3.2 that we maintain statistics $\overline{p_S}$ and $\overline{p_O}$ for each predicate $p \in P$, where $P$ is the set of all predicates in the data. Let $P_s$ and $P_o$ be the set of all $\overline{p_S}$ and $\overline{p_O}$, respectively. The following three conditions are checked for each predicate $p$: $(i)$ $p$ is the reserved *rdf:type* predicate; $(ii)$ $\overline{p_S}$ is three standard deviations[2] away from the arithmetic mean of all $\overline{p_S} \in P_s$; $(iii)$ $\overline{p_O}$ is three standard deviations away from the arithmetic mean of all $\overline{p_O} \in P_o$. The first condition prevents the type vertices from being selected as core vertices. Similarly, the other two conditions treat extremely high degree vertices as outliers and prevent them from being cores. If any of the previous conditions is satisfied, then we set: $\overline{p_S} = \overline{p_O} = -\infty$; else use the $\overline{p_S}$ and $\overline{p_O}$ as computed in Section 3.2. Now, we can compute a score for each vertex as follows:

DEFINITION 2 (VERTEX SCORE). *For a query vertex $v$: Let $E_{out}(v)$ be the set of outgoing edges (i.e., predicates where $v$ appears as subject), and $E_{in}(v)$ the set of incoming edges (i.e., predicates where $v$ appears as object). Also, let $S$ be the set of all $\overline{p_S}$ for the $E_{out}(v)$ edges and all $\overline{p_O}$ for $E_{in}(v)$ edges. The vertex score $\overline{v}$ is defined as: $\overline{v} = \max(S)$.*

Figure 7 shows an example. For vertex ?d, $E_{out}(?d) = \{\texttt{subOrgOf, type}\}$ and $E_{in}(?d) = \{\texttt{memberOf}\}$. The corresponding predicate scores are 3, 1 and 4. Therefore, $\overline{?d} = 4$.

DEFINITION 3 (CORE VERTEX). *Given a query $Q$, the vertex $v'$ with the highest score is called the core vertex.*

### 4.2 Generating the Redistribution Tree

Let $Q$ be a frequent query that PHD-Store decided to redistribute. Our goal is to generate a redistribution tree that minimizes the expected amount of replication. In Section 4.1 we explained why starting from the vertex with the highest score has the potential to minimize replication. Intuitively, the same idea applies recursively to each level of the redistribution. Therefore, our query redistribution tree spans all the edges of the query graph, such that every child node in the tree potentially has a lower (or equal) score than its parent. Each of the edges in the query graph should appear exactly once in the tree; vertices may be repeated.

Using the scoring function discussed in the previous section, we transform $Q$ into a vertex weighted, undirected graph $G$. The vertex with the highest score is selected as the core vertex. Then, $G$ is decomposed into the redistribution tree using Algorithm 2. The algorithm keeps exploring edges starting from high score vertices towards lower score ones. All edges incident to the core vertex $v'$ are inserted in a pending edges set. Then, the algorithm gradually keeps

---
[2]Huang et al. [20] also use the same cut-off threshold.

**Algorithm 2:** Generate Redistribution Tree

**Input:** $G = \{V, E\}$; a vertex-weighted, undirected query graph, $v'$; the core vertex
**Result:** A tree $T$

1 Let *core_edges* be all incident edges to $v'$;
2 **foreach** *edge e in core_edges* **do**
3     $e.parent \leftarrow \phi$;
4     add $e$ to *pendingList*;
5 **while** *pendingList notEmpty* **do**
    // e is the edge connected to the highest score vertex
6     $e \leftarrow$ getHighestScoreEdge(*pendingList*);
7     remove $e$ from *pendingList*;
    // extend the path of e.parent with e;
8
9     appendToPath($e$, $e.parent$);
10     $adj \leftarrow$ getAdjacentEdgesTo($e$);
11     **foreach** $\alpha$ *in adj* **do**
12         **if** $\alpha$ *NOT explored* **then**
13             $\alpha.parent \leftarrow e$;
14             add $\alpha$ to *pendingList*;
15 $T.root = v'$

exploring new edges by removing the edge with the highest vertex score first from the set, and inserting all its adjacent edges to the pending edges. Note that the direction of traversal of the graph is independent from the actual edge directions of the query. The result is a tree with the core vertex $v'$ as root. As an example, consider the query in Figure 7. Having the highest score, *?u* is chosen as core, and the query is decomposed into the tree shown in Figure 8. Note that the directions of the edges in Figure 8 are only used to define the join columns i.e., subject-subject, object-object or subject-object; they do not influence the tree traversal.

### 4.3 PHD: Propagating Hash Distribution

Our redistribution algorithm is a hybrid of hash partitioning and $k$-hop replication. Given a redistribution tree, PHD-Store distributes the data along paths from the root to leaves using breadth first traversal. The algorithm proceeds in two phases: First, it distributes triples that contain the core vertex to workers using a hash function $\mathcal{H}(\cdot)$. Let $t$ be such a triple and denote $t.v'$ as its core vertex (the core can be either the subject or the object of $t$). Let $w_1, w_2, \ldots, w_N$ be the workers. $t$ will be replicated in worker $w_j$, where: $j = \mathcal{H}(t.v') \mod N$.

In Figure 8, consider the first hop ⟨?d, $subOrgOf$, ?u⟩ of the highlighted path. Using the previous formula, the core *?u* determines the placement of $t_1$, $t_2$ and $t_3$ (see Table 1). Assuming two workers, $t_1$ and $t_2$ are replicated in $w_1$ (because of Stanford), whereas $t_3$ is replicated in $w_2$ (because of MIT). *?u* is called the *source* column of these triples.

DEFINITION 4 (SOURCE COLUMN). *The source column of a triple ⟨s, p, o⟩ is the column (subject or object) that is used to determine its placement.*

The second phase of PHD places triples of the remaining levels of the tree in the workers that contain their parent, through a series of distributed semi-joins. The column at the opposite end of the source column of the previous step becomes the *propagating* column; in our example, the propagating column is *?d*.

DEFINITION 5 (PROPAGATING COLUMN). *The propagating column of a triple ⟨s, p, o⟩ is the column (subject or object) that is at the opposite end of the corresponding source.*

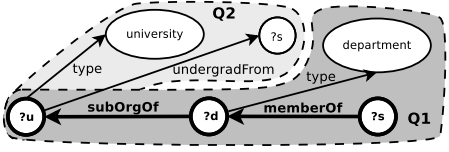

**Figure 8: Redistribution tree for the query in Figure 7. The selected (bold) part of the tree is a path from core vertex ?u to leaf ?s.**

|       | RDF triple                                              | Worker |
|-------|---------------------------------------------------------|--------|
| $t_1$ | ⟨Stanford-CS, $subOrgOf$, Stanford⟩                     | $w_1$  |
| $t_2$ | ⟨Stanford-ENG, $subOrgOf$, Stanford⟩                    | $w_1$  |
| $t_3$ | ⟨MIT-CS, $subOrgOf$, MIT⟩                               | $w_2$  |
| $t_4$ | ⟨Ben, $memberOf$, Stanford-CS⟩                          | $w_1$  |
| $t_5$ | ⟨Prof.James, $memberOf$, Stanford-ENG⟩                  | $w_1$  |
| $t_6$ | ⟨John, $memberOf$, Stanford-ENG⟩                        | $w_1$  |
| $t_7$ | ⟨Peter, $memberOf$, MIT-CS⟩                             | $w_2$  |

**Table 1: Triples from Figure 1 matching the highlighted path of Figure 8.**

*It is the join attribute that determines the placement of the next level of triples.*

In Figure 8 the second subquery of the highlighted path is ⟨?s, $memberOf$, ?d⟩. ?d from the previous level becomes now the source column. Triples $t_{4...7}$ (see Table 1) match the sub-query and are joined with triples $t_{1...3}$. Therefore, $t_4$, $t_5$ and $t_6$ are propagated to worker $w_1$, whereas $t_7$ is propagated to $w_2$. The process is formally described in Algorithm 3. The algorithm runs in parallel on all workers. Lines (5-8) perform hash distribution of all sub-queries incident to the core; we call this propagation level 0. Propagation to next levels is done through a series of semi-joins between each level in the path and the level directly before. Only triples that satisfy the join condition with the previous level are kept. This procedure causes triples on level $i$ to follow the placement of the triples in the parent level $i-1$ (lines 9-18).

Note that our PHD algorithm groups the data accessed by the query around the bindings of the core vertex, Stanford and MIT in this case. Redistributing more queries around the same core will allow future queries, that share that core, to execute in parallel mode by utilizing any combination of the previous redistributions. This cannot be achieved by executing queries using distributed semi-joins and materializing the intermediate results.

### 4.4 Proactive Redistribution

So far PHD-Store has been presented as a reactive system, where only frequent queries are redistributed and hence optimized for. Non-frequent queries, like the ones in Figure 9(a), that share the same pattern but have different cores (i.e Stanford, MIT and ?u) would be executed using the expensive distributed semi-join, even if they share a common frequent pattern. PHD-Store solves this problem by assigning such queries to the same query template and then redistributing the template.

DEFINITION 6 (QUERY TEMPLATE). *A query template $\mathcal{Q}$ is the query pattern that results from replacing all the constants in a query $Q$ with variables.*

Template vertices store the values of all matching query vertices and their counts. If a query matches an existing

**Algorithm 3**: Performing PHD on a given tree $T$ that resulted from the decomposition of query $Q$

**Input**: Query graph decomposed as a labeled tree $T$. $L$ is number of levels in $T$
**Result**: Data propagated from the root toward leaves

1 **Let** *source_nodes* be the set of source vertices;
2 **Let** *propagation_nodes* be the set of propagation vertices;
3 addToSourceNodes(*root*);
4 addToPrpagationNodes(*all children of root*);
   // hash-distributing (core-adjacent) edges
5 **foreach** *node v in propagation_nodes* **do**
      // subquery defined by *root*, *v* and the label of the edge between them
6     $sub_v \leftarrow$ constructSubQuery(*root, v, label*);
      // use the replica index to check
7     **if** $sub_v$ *is not previously distributed* **then**
8        hash-distribute all bindings of $sub_v$ on *root*;

9 **foreach** *level l: $2 \to L$* **do**
10     *source_nodes* $\leftarrow$ *propagation_nodes*;
11     clear *propagation_nodes*;
12     addToPrpagationNodes(*all children of source nodes*);
13     **foreach** *node v in propagation_nodes* **do**
14        $sub_v \leftarrow$ constructSubQuery(*v.parent, v, label*);
         // use the replica index to check
15        **if** $sub_v$ *is not previously distributed* **then**
16           $sub_{parent} \leftarrow$ constructSubQuery(*v.parent, v.grandParent, label*);
            // Join $sub_v$ and $sub_{parent}$ using distributed semi-join
17           *qualified* $\leftarrow sub_v \bowtie sub_{parent}$;
18           addDataToIndex(*qualified*);

template, the template vertices are updated to count their matched values, otherwise, a new template is created and initialized. For example, assume queries $Q_1$, $Q_2$, and $Q_3$ in Figure 9(a) were executed in this order. After executing $Q_1$, the template in Figure 9(b) will be created with vertex values shown in $T_1$. Template vertex values shown in $T_2$ reflect the template state after executing all three queries.

The frequency of a query template is the number of queries mapped to it. Once the template's frequency increases above a system-wide threshold, the redistribution process is triggered. Redistributing a template results in a replicated index that solves future matching queries efficiently. Nevertheless, redistributing a template involves shuffling large amounts of data and results in significant replication since all template vertices are variables. PHD-Store introduces a proactivity threshold that is used to balance the expected performance gain and the excessive cost of redistributing a template. If the number of unique values in a template vertex is greater than the proactivity threshold, the vertex is kept as a variable. Otherwise, only the most frequent value is assigned to the template vertex. In our example, assuming proactivity threshold of 2, vertices $V_2$ and $V_3$ in Figure 9(b) are replaced by ?d and *dept*, respectively.

### 5. REPLICA AND QUERY INDEX

**Replica index.** Each worker has a replica index consisting of a set of trees that have been redistributed, together with the replicated data. Consider the following queries:

```
Q1:SELECT ?u WHERE {        Q2:SELECT ?s WHERE {
  ?d subOrgOf  ?u             ?s  undergradFrom  ?u
  ?d type   department        ?u  type   university
  ?s memberOf  ?d           }
}
```

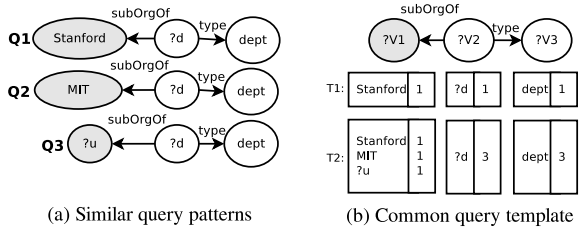
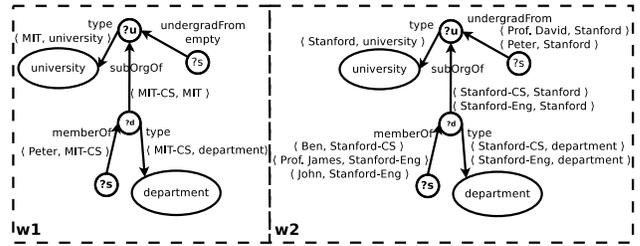

| (a) Similar query patterns | (b) Common query template |

Figure 9: Similar queries in (a) are assigned to the query template in (b).

Figure 10: The state of the replica index on two workers after redistributing $Q_1$ and $Q_2$.

Assume $Q_1$ is the first query to trigger redistribution, so the replica index is empty, and let $?u$ be the core. The corresponding redistribution plan is labeled $Q_1$ in Figure 8. Triples that bind to $\langle\text{?d}, subOrgOf, \text{?u}\rangle$ will be hash distributed on $?u$. A new tree is created in the replica index of every worker with $?u$ as root. Then, a vertex $?d$ is created with $?u$ as its parent and edge $(?d, ?u)$ is labeled with subOrgOf. Moreover, a storage module (see Section 3.1) is created and associated with this edge; it stores the data that are hashed to the respective worker. Using the binding values of $?d$, propagation continues by fetching the triples that bind to $\langle\text{?d}, type, \text{department}\rangle$. A new vertex $department$ is created as a child of $?d$ in all workers; and edge $(?d, department)$ is labled with type. The storage module associated with that edge stores the data that are hashed to the corresponding worker. The process continues recursively for $\langle\text{?s}, memberOf, \text{?d}\rangle$. $Q_2$ is redistributed in the same manner by extending the previous tree. Figure 10 shows the state of the replica index on two workers after redistributing $Q_1$ and $Q_2$ for the RDF graph in Figure 1.

Recall that each worker stores the original data in its main index. There are three reasons for storing the redistributed data in the separate replica index: $(i)$ as more queries are redistributed, updating a single index becomes a bottleneck; $(ii)$ because of replication, using one index mandates filtering duplicate results like in [20]; and $(iii)$ if data is coupled in a single index, intermediate join results will be larger, which will affect performance.

**Query index.** The query index is only created and maintained by the master. It has exactly the same structure as the replica indices of the workers, but the query index does not include storage modules and does not store any data. Instead, it is used by the query planner to check if a query can be executed in parallel mode. When a new query $Q$ is posed, the planner decomposes $Q$ into its redistribution tree $\tau$. If $\tau$ shares the same root with a tree in the the query index and all of $\tau$'s edges exist in the query index, then $Q$ can be answered in parallel mode; otherwise, $Q$ is answered using distributed semi-joins. For example, as a result of redistributing $Q_1$ and $Q_2$, the entire query in Figure 8 can be executed in parallel mode since its redistribution tree exists in the query index.

**Conflicting redistributions.** Conflicts arise when a subquery appears at two different levels in the query index. For example, suppose that after redistributing $Q_1$ and $Q_2$, we want to redistribute $Q_3 = \langle\text{?d}, subOrgOf, \text{?u}\rangle$ AND $\langle\text{?p}, worksFor, \text{?d}\rangle$, and let $?p$ be the core of $Q_3$. An edge associated with $\langle\text{?d}, subOrgOf, \text{?u}\rangle$ will be created in the second level of the query and replica indices. Recall that a similar edge was created in the first level because of $Q_1$. This may cause some triples to be replicated in two levels. In terms of correctness, this is not a problem for PHD-Store, because conflicting triples (if any) are stored separately using two different storage modules. To answer $Q_3$ in parallel mode, the subquery $\langle\text{?d}, subOrgOf, \text{?u}\rangle$ is answered using the data stored in the second level. This approach avoids the burden of any house keeping and duplicates management. The trade-off, however, is more memory consumption. The system currently has a parameter $\varrho_{max}$ that the administrator can set to enforce a *maximum replication ratio*. PHD-Store will redistribute queries as long as the replication ratio is lower than $\varrho_{max}$.

**Unbounded predicates.** Queries with unbounded predicates, such as $?p$ in $\langle\text{?x}, ?p, \text{Prof.David}\rangle$, pose a challenge when it comes to data locality. For instance, long path queries with unbounded predicates require replication of multiple hops to ensure parallel execution. In this case, replication grows exponentially [20] and would consume the memory. PHD-Store can still answer such queries using distributed semi-joins. However, we redistribute only queries with bounded predicates.

## 6. UPDATES

PHD-Store supports efficient batch updates in the form of insertion and deletion of RDF triples. The process is explained below. Note that the structure of the query and replica indices never changes because of update operations. Only the main index and the data in the storage modules of the replica index are affected.

**Deletion.** Delete operations can run in parallel mode without communication among workers. The triples to be removed are sent to each worker, which checks if any of the triples exists in its main index and deletes it accordingly. Note that a triple $t$ may exist in the main index of one worker only, but it may appear in many replica indices. Therefore, each worker traverses its replica index using depth-first search and deletes any instances of $t$. It also removes from the replica index all triples that are associated with $t$. Suppose we want to delete $t_1 = \langle\text{MIT-CS}, subOrgOf, \text{MIT}\rangle$ and $t_2 = \langle\text{MIT}, type, \text{university}\rangle$ from the replica indices shown in Figure 10. Each worker traverses its replica index to remove these triples. $t_1$ only exists in worker $w_1$, stored in the subOrgOf edge. After deleting $t_1$, there will be no other triples that share value MIT-CS for variable $?d$. Therefore, the triples that have MIT-CS as a binding of $?d$ are removed as well; these are $\langle\text{Peter}, memberOf, \text{MIT-CS}\rangle$ at edge memberOf and $\langle\text{MIT-CS}, type, \text{department}\rangle$ at edge type. The process continues recursively to the next levels. The same process applies to $t_2$, but it will not cause any additional cleanup because university is a leaf node. The replica index of worker $w_2$ is not affected. Figure 11(a) shows the state of the replica index of $w_1$ after deleting $t_1$ and $t_2$.

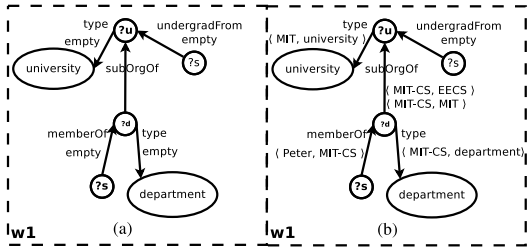

**Figure 11:** On worker 1: (a) The replica index after deletions. (b) The replica index after insertions.

| | LUBM | | YAGO2 | |
|---|---|---|---|---|
| | PHD | 2-hop | PHD | 2-hop |
| Preprocessing | 0 | 47,700 | 0 | 78,372 |
| Loading & indexing | 65 | 176 | 73 | 249 |
| Statistics collection | 6 | 0 | 7 | 0 |
| TOTAL | 71 | 47,876 | 80 | 78,621 |

**Table 2:** Startup time (sec); LUBM and YAGO2

**Insertion.** To insert a new triple $t$, the master first assigns it randomly to a worker, which inserts $t$ in its main index. Next, each worker traverses its replica index to ensure consistency. Continuing our example, assume that after the previous deletions have been executed, we want to insert three triples: $t_3 = \langle$MIT-CS, $subOrgOf$, EECS$\rangle$, $t_4 = \langle$MIT, $type$, university$\rangle$ and $t_5 = \langle$MIT-CS, $subOrgOf$, MIT$\rangle$. Assume that value EECS hashes to worker $w_1$. Then, $t_3$ is inserted at the subOrgOf edge in the replica index of $w_1$. At that edge, no other triples share with $t_3$ the same value MIT-CS for the ?d vertex. Therefore, the subtree rooted at ?d has to be validated and made consistent. This is done by sending requests to all other workers. As a result, $\langle$Peter, $memberOf$, MIT-CS$\rangle$ and $\langle$MIT-CS, $type$, department$\rangle$ are propagated to $w_1$ and stored at edges memberOf and type, respectively. Similarly, assuming that MIT hashes to $w_1$, $t_4$ is inserted in the ?u–type–university edge of the replica index. No validation is required because university is a leaf. Triple $t_5$ is inserted in $w_1$ in the same way as $t_3$. Note that, because $t_5$ shares the same value MIT-CS with $t_3$, no validation is required since it was already done after inserting $t_3$. Figure 11(b) shows the state of the replica index of worker $w_1$ after executing all insertions.

## 7. EXPERIMENTAL EVALUATION

We implemented: (*i*) PHD-Store in C++ using MPI for synchronization and communication; and (*ii*) *SemiJoin*, a baseline approach that uses distributed semi-joins to execute queries, without redistribution or replication. Both systems work on randomly partitioned data. Our closest competitor is the *k*-hop system by Huang et al. [20]. For *k*-hop, we partitioned the graph using the parallel version of METIS [22], and used Hadoop and the code provided by the authors for triple placement. For fairness, instead of a slow disk-based store, we implemented an in-memory query engine for *k*-hop using the same data structures as PHD-Store. We used the 2-hop configuration from Huang et al. as it performs better than hash partitioning and 1-hop guarantee [20].

We used two popular datasets: the synthetic LUBM [1] benchmark and the real YAGO2 dataset [19]. For LUBM, we generated a dataset of 2,000 universities, resulting in almost 50GB of raw data or 267M triples. We used 12 queries[3] from the benchmark that contain at least one join. We also added one complex query ($QS$) that is more than 2 hops long. From these 13 queries, we generated 12K similar ones that have the same patterns but different constants. We constructed 5 workloads. Each workload consists of 20K queries which are randomly selected from the 12K queries. YAGO2 is a real dataset derived from Wikipedia, WordNet and GeoNames. From the native YAGO2 format we extracted around 30GB of raw data, or around 300M triples. We generated a sequence of 1,000 queries randomly from queries A1, A2, B1, B2 and B3 defined in Binna et al. [7]. All queries are available online[4].

---
[3]Only patterns are used; inference is outside our scope.

We used a cluster with 21 machines; one is the master and the remaining 20 are workers. Each machine has an Intel i5-660 3.3GHz CPU (dual core), 16GB RAM and 2x2TB (7200 RPM, 6.0Gb/s) hard disks in RAID-0 configuration. The machines run 64-bit 2.6.38-8 Linux Kernel and are connected via a 1GBps Juniper switch. We restricted the bandwidth of the switch to 10MBps to simulate typical WAN speeds.

### 7.1 Startup Time

This experiment measures the time it takes PHD-Store and $k$-hop to prepare the data prior to answering queries. The results are shown in Tables 2 for LUBM and YAGO2. 2-hop spends a lot of time preprocessing the data, the most expensive step being graph partitioning. Data loading and indexing also take more time in 2-hop because of the existence of replicas. PHD-Store, on the other hand, incurs some cost for statistics collection. In total, PHD-Store needs almost 3 orders of magnitude less time than its competitor. PHD-Store can start answering queries in around 1 minute, whereas 2-hop needs up to 22 hours.

For fair comparison, we reconfigured Hadoop so intermediate HDFS files are written into a memory mounted partition. Then, we re-evaluated the preprocessing phase of 2-hop. As expected, the preprocessing time dropped to 25,508 and 28,115 seconds for LUBM and YAGO2, respectively. Nonetheless, the startup cost for PHD-Store is still 500 times less than 2-hop.

### 7.2 Frequency and Proactivity Thresholds

The frequency and proactivity thresholds control the triggering of redistributions. They are highly correlated and both influence the execution time and the amount of replication. In this experiment, we select the thresholds values based on one of the random LUBM workloads. We first set the proactivity threshold to infinity and execute the workload by varying the frequency threshold values. The execution time and the resulting replication ratio are shown in Figures 12(a) and 12(b), respectively. As the frequency threshold increases, the execution time increases as most of the queries use expensive semi-joins. At the same time, the higher the frequency threshold, the lower the replication ratio because fewer queries are redistributed. For this reason, we select a frequency threshold of 3 as the default value in all experiments. Next, we vary the proactivity threshold. Based on the results shown in Figures 12(c) and 12(d), we set the proactivity threshold to 10 for all the experiments.

---
[4]http://cloud.kaust.edu.sa/Pages/queries.aspx

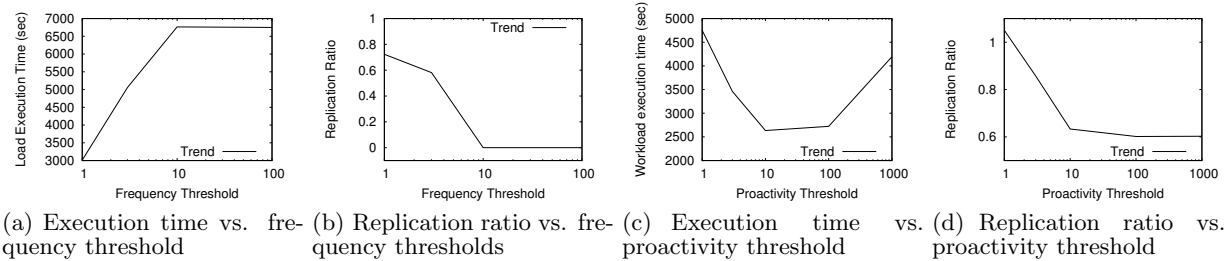

Figure 12: Sensitivity analysis for frequency and proactivity thresholds.

### 7.3 Query Execution Cost

**Entire workload.** Next we measure the cumulative execution time during the execution of an entire workload. The cumulative time includes the preprocessing cost for $k$-hop and the cost of dynamic redistribution for PHD-Store. Figure 13(a) shows the average cumulative time for the 5 random query loads generated from the 12K queries. $k$-hop pays most of the cost at the preprocessing phase. PHD-Store, on the other hand, amortizes the redistribution cost during the actual query processing; the spikes in the graph are due to redistribution. Note that the proactive version of PHD-Store is at least 2 times faster than the reactive version. After roughly 10,000 queries the system converges. In all cases PHD-Store is 1 to 2 orders of magnitude faster than $k$-hop. Figure 14(a) shows the results for YAGO2. Because of the limited number of queries, both reactive and proactive PHD-Store have the same performance. Again, PHD-Store is 2 orders of magnitude faster than $k$-hop.

**Single queries.** The next experiment measures the effectiveness of the partitioning. We allow all systems to converge (i.e., preprocessing and dynamic redistribution costs are excluded); then we measure the execution time for each of the 13 LUBM queries. The results are shown in Figure 13(b). PHD-Store achieves significantly better (see queries 2 and QS), or comparable performance to 2-hop. The only exception is query 8, where 2-hop is 10msec faster. The radius of query 8 is at most 2, therefore 2-hop executes it in parallel mode using all workers. PHD-Store, on the other hand, selects a constant (i.e., `university0`) as core vertex. Therefore the redistribution process sends all relevant triples to the same worker. The query is still answered without communication overhead, but there is no parallelism, resulting in slower execution. SemiJoin, the baseline method that does not redistribute any data, is 1 to 3 orders of magnitude slower than PHD-Store. This demonstrates the significance of good data placement.

Figure 14(b) shows the results for the YAGO2 dataset. Most of the vertices in the query graphs are not connected to an $rdf$:$type$ predicate. Without this information, 2-hop cannot distinguish whether it can process a certain query in parallel mode, even if the radius of the query is indeed at most 2. Therefore, most queries are executed by expensive distributed semi-joins. PHD-Store is 1 to 2 orders of magnitude faster for all queries.

### 7.4 Replication Cost

Figure 13(c) shows the average net replication ratio as a percentage of the original data size, after completing the 5 LUBM workloads. The LUBM workloads contain queries with long chains (more than 2 hops in radius), which are very costly when executed by 2-hop; and require the extension to 3-hop to be answered efficiently. In contrast, by only redistributing what is needed PHD-Store performs better while incurring less replication. Proactive PHD-Store results in more replication than the reactive version as non-frequent queries may be redistributed. For YAGO2, the net replication of PHD-Store is lower than that of 2-hop. Figure 14(c), shows that 2-hop results in replication that is almost 2 times the original data.

Recall from Section 5 that $\varrho_{max}$ limits the maximum replication ratio. Figure 15(a) shows the effect of varying $\varrho_{max}$ on the execution time, for an LUBM workload. Constraining the replication to low values results in blocking query redistributions. Accordingly, many frequent queries run using expensive distributed semi-joins. On the other hand, when $\varrho_{max}$ increases, many of the frequent queries are redistributed and hence run efficiently in parallel mode.

### 7.5 Workload Coverage

PHD-Store does not redistribute all parts of the data graph, therefore and for fair comparison, we examine the influence of the workload coverage on the performance of the $k$-hop system. We extracted only triples that are relevant to one of the 5 LUBM workloads. Then, we partitioned this data using 2-hop guarantee. Afterwards, we executed the same workload on the partitioned data. The query load we selected covers around 61% of the original data. As shown in Table 7.5, it took $k$-hop 11.7 hours to partition the covered data; roughly 1.5 hours less than partitioning the whole dataset. The wall time for executing the workload (excluding preprocessing) dropped by 21%. The gain in query performance is masked by the expensive preprocessing overhead. Moreover, such a partitioning is only useful for answering this specific workload. Repartitioning the entire dataset is needed for different query loads.

### 7.6 Scalability and Load Balancing

In the following experiment, we measure the scalability of PHD-Store, using the LUBM workloads. We used the LUBM generator to generate datasets of different sizes. We vary the number of universities from 500 to 2,000; the resulting datasets contain from 67M to 267M triples. Figure 15(b)

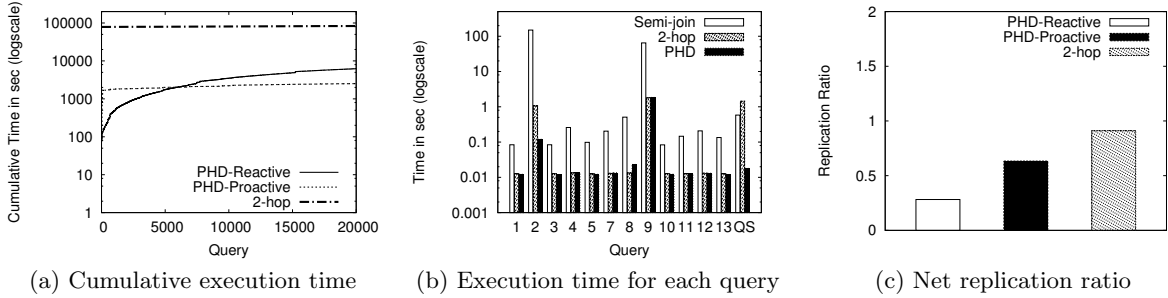

(a) Cumulative execution time  (b) Execution time for each query  (c) Net replication ratio

Figure 13: Execution time and replication cost for LUBM.

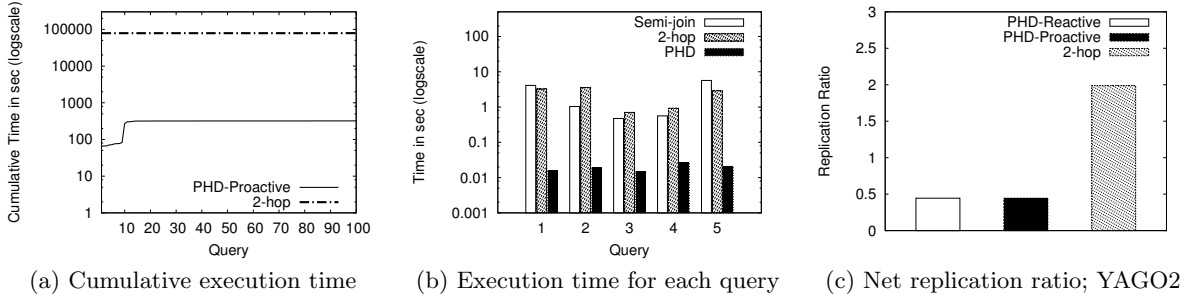

(a) Cumulative execution time  (b) Execution time for each query  (c) Net replication ratio; YAGO2

Figure 14: Execution time and replication cost for YAGO2.

|  | Partial | Full |
|---|---|---|
| Preprocessing (sec) | 42,156 | 47,700 |
| Load execution (sec) | 3,816 | 4,831 |
| TOTAL | 45,972 | 52,531 |

Table 3: Workload coverage effect

|  | Delete | Insert |
|---|---|---|
| Throughput (triples/sec) | 58,886 | 28,485 |

Table 4: PHD-Store update operation throughput

shows that scalability is linear, demonstrating that PHD-Store can scale to very large datasets.

In the final experiment, we measure how balanced the data placement is after redistribution. We use the Gini coefficient [16], whose values range from 0 (i.e., perfect balance) to 1. Large values of the coefficient imply that some workers will be overwhelmed in terms of storage requirements and computational time during query processing. Figure 15(c) shows the average gini coefficient after executing the LUBM and YAGO2 workloads. PHD-Store achieves balanced distribution. For LUBM, the Gini coefficient of PHD-Store is almost an order of magnitude lower than 2-hop. Similarly, for YAGO2, PHD-Store's Gini coefficient is at least 2 orders of magnitude better than 2-hop.

### 7.7 Updates

In this experiment, we show the performance of PHD-Store when executing update operations. As discussed in Section 5, updating the main index poses no overhead on the system, however, maintaining replica consistency can be overwhelming. Therefore, we should execute update queries after executing a whole query workload. To evaluate the delete operation, we allow the system to execute one of the LUBM workloads and redistributes queries from that workload. Then, we delete 20% of the data (≈ 53.4M triples) and measure the deletion time. The insert operation is simply evaluated by re-inserting the deleted triples into the system. Deletions and insertions were carried out in batches. Each batch consists of ≈ 1.1M triples. Table 7.7 shows the throughput of our system when executing updates. As expected, deletions are more efficient than insertions as they do not require communication among workers.

### 8. CONCLUSION

This paper presented PHD-Store, a SPARQL engine for federations of many independent RDF repositories. As more organizations publish their data in the versatile RDF format, rendering the deep web accessible to search engines and end users, the demand for federating RDF repositories is expected to increase. Without proper data placement, executing queries against such federations is very costly. PHD-Store follows an adaptive approach that allows it to start processing queries immediately, thus minimizing the data-to-query-time, while it dynamically distributes and indexes only those parts of the graph that benefit the most frequent query patterns. The experimental results verify that PHD-Store achieves better partitioning and replicates less data than its competitors. More importantly, PHD-Store scales to very large RDF graphs, whereas existing methods are limited to much smaller datasets by prohibitively expensive preprocessing. Currently we are working on a disk-based version of our system in order to support even larger datasets. We are also investigating the possibility of utilizing PHD-Store for general (i.e., non-RDF) graphs, and operators such as graph traversals, or reachability queries.

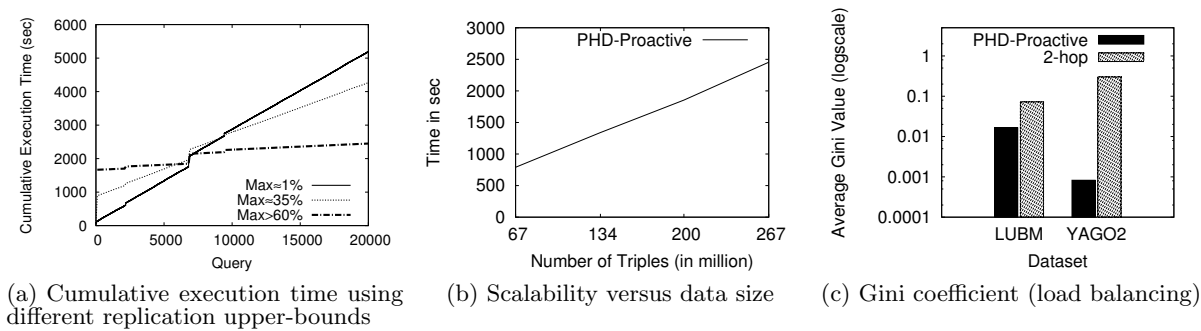

Figure 15: Cumulative execution time, scalability and load balancing; LUBM workload.

## 9. REFERENCES

[1] LUBM SPARQL Benchmark. http://swat.cse.lehigh.edu/projects/lubm/.
[2] NuoDB. http://www.nuodb.com/.
[3] RDF Primer. http://www.w3.org/TR/rdf-primer/.
[4] D. Abadi, A. Marcus, S. Madden, and K. Hollenbach. Scalable semantic web data management using vertical partitioning. In *VLDB*, 2007.
[5] I. Alagiannis, R. Borovica, M. Branco, S. Idreos, and A. Ailamaki. NoDB in action: adaptive query processing on raw data. *PVLDB*, 5(12), 2012.
[6] S. Alexaki, V. Christophides, G. Karvounarakis, D. Plexousakis, and K. Tolle. The ICS-FORTH RDFSuite: Managing Voluminous RDF Description Bases. In *SemWeb*, 2001.
[7] R. Binna, W. Gassler, E. Zangerle, D. Pacher, and G. Specht. SpiderStore: A Native Main Memory Approach for Graph Storage. In *Grundlagen von Datenbanken*, volume 733, 2011.
[8] S. Blanas, Y. Li, and J. M. Patel. Design and evaluation of main memory hash join algorithms for multi-core CPUs. SIGMOD, 2011.
[9] M. A. Bornea, J. Dolby, A. Kementsietsidis, K. Srinivas, P. Dantressangle, O. Udrea, and B. Bhattacharjee. Building an efficient RDF store over a relational database. SIGMOD, 2013.
[10] J. Broekstra, A. Kampman, and F. V. Harmelen. Sesame: A Generic Architecture for Storing and Querying RDF and RDF Schema. In *ISWC*, 2002.
[11] R. Castillo and U. Leser. Selecting materialized views for RDF data. ICWE, 2010.
[12] Z. Chong, H. Chen, Z. Zhang, H. Shu, G. Qi, and A. Zhou. RDF pattern matching using sortable views. CIKM, 2012.
[13] C. Curino, E. Jones, Y. Zhang, and S. Madden. Schism: a workload-driven approach to database replication and partitioning. *PVLDB*, 3(1-2), 2010.
[14] V. Dritsou, P. Constantopoulos, A. Deligiannakis, and Y. Kotidis. Optimizing query shortcuts in RDF databases. ESWC, 2011.
[15] O. Erling. Towards Web Scale RDF. In *SSWS*, 2008.
[16] C. Gini. Concentration and Dependency Ratios (1909, in Italian). English translation in Rivista di Politica Economica, 87, 1997.
[17] F. Goasdoué, K. Karanasos, J. Leblay, and I. Manolescu. View selection in Semantic Web databases. *PVLDB*, 5(2), 2011.
[18] A. Harth, J. Umbrich, A. Hogan, and S. Decker. YARS2: A Federated Repository for Querying Graph Structured Data from the Web. In *ISWC/ASWC*, volume 4825, 2007.
[19] J. Hoffart, F. Suchanek, K. Berberich, E. Lewis-Kelham, G. de Melo, and G. Weikum. YAGO2: exploring and querying world knowledge in time, space, context, and many languages. In *Proc. WWW*, 2011.
[20] J. Huang, D. Abadi, and K. Ren. Scalable SPARQL Querying of Large RDF Graphs. *PVLDB*, 4(11), 2011.
[21] S. Idreos, M. L. Kersten, and S. Manegold. Database Cracking. In *CIDR*, 2007.
[22] G. Karypis and V. Kumar. MeTis: Unstructured Graph Partitioning and Sparse Matrix Ordering System. http://www.cs.umn.edu/~metis, 2009.
[23] B. Liu and B. Hu. HPRD: a high performance RDF database. *Int. J. Parallel Emerg. Distrib. Syst.*, 2010.
[24] G. Malewicz, M. Austern, A. Bik, J. Dehnert, I. Horn, N. Leiser, and G. Czajkowski. Pregel: a System for Large-scale Graph Processing. In *SIGMOD*, 2010.
[25] T. Neumann and G. Weikum. RDF-3X: a RISC-style engine for RDF. *PVLDB*, 1(1), 2008.
[26] A. Pavlo, C. Curino, and S. Zdonik. Skew-Aware Automatic Database Partitioning in Shared-Nothing, Parallel OLTP Systems. In *SIGMOD*, 2012.
[27] K. Rohloff and R. E. Schantz. High-performance, massively scalable distributed systems using the MapReduce software framework: the SHARD triple-store. In *Programming Support Innovations for Emerging Distributed Applications*, 2010.
[28] M. Stonebraker, S. Madden, D. Abadi, S. Harizopoulos, N. Hachem, and P. Helland. The end of an Architectural Era: (It's Time for a Complete Rewrite). In *VLDB*, 2007.
[29] C. Weiss, P. Karras, and A. Bernstein. Hexastore: sextuple indexing for semantic web data management. *PVLDB*, 1(1), 2008.
[30] K. Wilkinson, C. Sayers, H. Kuno, and D. Reynolds. Efficient RDF Storage and Retrieval in Jena2. In *SWDB*, pages 131–150, 2003.
[31] S. Yang, X. Yan, B. Zong, and A. Khan. Towards effective partition management for large graphs. In *SIGMOD*, 2012.